\documentstyle[epsf]{cupconf}

\def\eps@scaling{.95}
\def\epsscale#1{\gdef\eps@scaling{#1}}
\def\plotone#1{\centering \leavevmode
\epsfxsize=\eps@scaling\textwidth \epsfbox{#1}}

\ifoldfss
\else
  \ifnfssone
    \newmathalphabet{\mathit}
      \addtoversion{normal}{\mathit}{cmr}{m}{it}
      \addtoversion{bold}{\mathit}{cmr}{bx}{it}
    \newmathalphabet{\mathcal}
      \addtoversion{normal}{\mathcal}{cmsy}{m}{n}
    \else
    \ifnfsstwo
    \fi
  \fi
\fi

\def\hexnumber#1{\ifcase#1 0\or1\or2\or3\or4\or5\or6\or7\or8\or9\or
 A\or B\or C\or D\or E\or F\fi }

\makeatletter
\ifx\CUP@mtlplain@loaded\undefined
\else
\fi
\makeatother
 \makeatletter
 \ifx\CUP@mtlplain@loaded\undefined
   \font\tenbmi=cmmib10 at 10pt
   \font\sevenbmi=cmmib10 at 7pt
   \font\fivebmi=cmmib10 at 5pt

   \newfam\bmifam
   \textfont\bmifam=\tenbmi
   \scriptfont\bmifam=\sevenbmi
   \scriptscriptfont\bmifam=\fivebmi
   
 \fi
 \makeatother
\ifnfsstwo

\fi
\ifnfssone

\fi
\ifoldfss

\fi

\mathchardef\varLambda="0103

\makeatletter
\ifx\CUP@mtlplain@loaded\undefined
\else
\fi
\makeatother
\makeatletter
\ifx\CUP@mtlplain@loaded\undefined
  \font\tenbms=cmbsy10
  \font\sevenbms=cmbsy10 at 7pt
  \font\fivebms=cmbsy10 at 5pt
  \newfam\bmsfam
  \textfont\bmsfam=\tenbms
  \scriptfont\bmsfam=\sevenbms
  \scriptscriptfont\bmsfam=\fivebms

  \edef\bsy@{\hexnumber\bmsfam}
  \mathchardef\bnabla="0\bsy@72
\fi
\makeatother
\def\ha{H$\alpha$}
\def\sii{[S~{\sc ii}]}
\def\hii{H~{\sc ii}}

\begin{document}
\ifnfssone
\else
  \ifnfsstwo
  \else
    \ifoldfss
      \let\mathcal\cal
      \let\mathrm\rm
      \let\mathsf\sf
    \fi
  \fi
\fi

  \title[Cambridge University Press]{Super-luminous
    Supernova Remnants}

  \author[Y.-H. Chu {\it et al.\/}]{%
  Y\ls O\ls U\ls -\ls H\ls U\ls A\ns 
  C\ls H\ls U,\ns 
  C.-H.\ns R\ls O\ls S\ls I\ls E\ns
  C\ls H\ls E\ls N\ns \\
  \and\ns
  S\ls H\ls I\ls H\ls -\ls P\ls I\ls N\ls G\ns 
  L\ls A\ls I}
  \affiliation{Astronomy Department, University of Illinois,
               1002 W. Green St., Urbana, IL 61801, USA}
  \maketitle

\begin{abstract}

Some extragalactic SNRs are more than two orders of magnitude
more luminous than the young Galactic SNR Cas A.  These SNRs 
are called super-luminous or ultra-luminous SNRs.  Their high
luminosities can be caused by chance superpositions of multiple
objects, interactions with a very dense environment, or unusually
powerful supernova explosions.  Four super-luminous SNRs are 
known: one in NGC\,4449, one in NGC\,6946, and two in M101.  
The two remnants in M101, NGC\,5471B and MF83, are recently 
suggested to be ``hypernova remnants" possibly connected to the 
GRBs.  We have obtained new or archival HST WFPC2 images and new 
high-dispersion echelle spectra of these super-luminous SNRs, in 
order to examine their stellar and interstellar environments and 
to analyze their energetics.  We discuss the physical nature of 
these four SNRs, with a special emphasis on the two ``hypernova 
remnants'' in M101.

\end{abstract}

\firstsection % if your document starts with a section,
              % remove some space above using this command.
\section{Introduction}

Supernova remnants (SNRs) usually exhibit three distinguishing 
characteristics: nonthermal radio emission, bright X-ray emission, 
and large optical \sii/\ha\ line ratios.  These characteristics
have been used as identification criteria for SNRs.  In the 
Galaxy, SNRs are most effectively surveyed at radio wavelengths 
because heavy extinction in the Galactic plane hampers optical and
X-ray observations (e.g., Green 1988).  In the Magellanic Clouds, 
where interstellar extinction is small, SNRs can be surveyed at 
radio, X-ray, or optical wavelengths (Mathewson et al.\ 1983, 1984, 
1985; Smith 1999).  In distant galaxies, the sensitivity of available 
radio and X-ray instruments becomes a limiting factor; thus, surveys
of SNRs are carried out in the optical (e.g., D'Odorico, Dopita, \& 
Benvenuti 1980; Blair \& Long 1997; Matonick \& Fesen 1997).  

Only a small number of the optically identified extragalactic SNR
candidates can be confirmed at radio or X-ray wavelengths.  These 
confirmed SNRs are among the most luminous remnants.  In fact, some 
of them are so luminous that they were first identified in radio or 
X-rays and subsequently confirmed by optical spectroscopic 
observations.  These luminous SNRs are often more than two orders 
of magnitude more luminous than the young Galactic SNR Cas A, and 
hence they have been called super-luminous or ultra-luminous SNRs.

Several factors may contribute to the brightness of extragalactic 
super-luminous SNRs.  First, they might be composite objects, 
with X-ray binaries contributing to the X-ray emission, multiple 
SNRs contributing to the radio emission, or \hii\ regions 
contributing to the \ha\ emission.  Second, these SNRs might be 
interacting with a very dense circumstellar or interstellar 
environment.  As optical and X-ray emissions are dependent on 
(density)$^2\times$(volume) and the energy is dependent on 
(density)$\times$(volume), for the same amount of energy, a greater 
amount of emission can be generated if the density is high.  
Finally, unusually powerful supernova explosions produce more 
energetic SNRs which may generate more emission and become 
super-luminous SNRs.  For example, two super-luminous SNRs in 
M101 have been suggested to be ``hypernova remnants" that require 
explosion energies as large as 10$^{53}$--10$^{54}$ ergs (Wang 1999).

In this paper, we report our analysis of four super-luminous 
SNRs.  Two of these four remnants are the ''hypernova remnants" 
in M101, and they are the main emphasis of this paper.  We have 
obtained high-resolution images to examine the nebular morphology
and stellar content, and high-dispersion spectra to determine the
expansion velocity and kinetic energy.  These new observations 
allow us to investigate the emission mechanisms and whether 
unusually energetic supernova explosions are needed for these
remnants.

\section{Super-luminous Supernova Remnants}

Four super-luminous SNRs have been reported: one in NGC\,4449
(Kirshner and Blair 1980), one in NGC6946 (Schlegel 1994;
Blair \& Fesen 1994; Van Dyk et al.\ 1994), and two in M101 
(Matonick \& Fesen 1998; Wang 1999).  The X-ray, radio, and 
optical H$\alpha$ luminosities of these four remnants and
Cas A (for comparison) are listed in Table 1.  The numbers
in Table 1 are taken from the above cited references and this 
paper.  The radio luminosity at 408 MHz has been normalized 
to Cas A for easy comparisons.

It is clear from Table 1 that these super-luminous SNRs are
most remarkable in the X-ray.  The SNR in NGC\,4449 is the
smallest, the SNR in NGC\,6946 has the highest X-ray luminosity,
NGC\,5471B has the highest H$\alpha$ luminosity, and MF83 is the
largest.

\begin{table}
%  \begin{center}
%    \begin{minipage}{6cm}
       \begin{tabular}{lccccc}
SNRs  & Distance  &    L(X)          &   L(408 MHz)   &  L(H$\alpha$)     &  Size  \\
      &  (Mpc)    &(10$^{36}$ erg/s) &   (Cas A)      & (10$^{37}$ erg/s) & (pc) \\
\\
Cas A     &  0.003 &    ~~~3    &       ~1       &   ---           &    4   \\
NGC\,4449 &  5~~  & ~800       &       25       &   $<$~0.15       & $<$ 0.6 \\
NGC\,6946 &  5.1  & 2800       &       ~1       &    ~19          & 20$\times$30 \\
NGC\,5471B&  7.2  & ~170       &       ~3       &    160          &   60  \\
MF83      &  7.2  & ~100       &       ---      &    ~17          &   267 \\
\\
        \end{tabular}
%     \end{minipage}
%   \end{center}
\caption{Super-luminous Supernova Remnants}
\end{table}

\section{Observations and Datasets}

We intend to determine the physical nature of these super-luminous
SNRs and the causes of their high luminosities.  Useful optical
observations for this study include high-resolution images and 
high-dispersion spectra.  High-resolution images in the continuum
bands can be used to study the stellar environment of a remnant,
while flux-calibrated H$\alpha$ images can be used to determine the 
H$\alpha$ luminosity and rms electron density of the remnant. 
The [S II]/H$\alpha$ ratio map can be used to examine the spatial
extent of the SNR shocks.  High-dispersion spectra are useful in 
separating shocked and unshocked components, determining expansion 
and shock velocities, and measuring diagnostic nebular line ratios. 

We have used the following datasets for 
the analysis presented in this paper: \\
\indent (1) HST FOC [O III] image of the SNR in NGC\,4449 (archival);  \\
\indent (2) HST WFPC2 H$\alpha$, [S II], and continuum images of the SNR in
    NGC\,6946 (archival),\\
\indent\indent ~~NGC\,5471B (proprietary); \\
\indent (3) MDM 2.4m B, V, I, and H$\alpha$ images of MF83 (courtesy of Eva 
           Grebel); and\\
\indent (4) KPNO 4m echelle spectra of the SNR in NGC\,6946, NGC\,5471B, and MF83 \\
  \indent\indent ~~(proprietary).

We have adopted the analysis of the super-luminous SNR in NGC\,4449 
by Blair \& Fesen (1998), but carried out our own analysis for the
other three super-luminous SNRs.  The detailed analysis and final results will be 
reported later in refereed journals.

\section{The Super-luminous SNR in NGC\,4449}

The super-luminous SNR in NGC\,4449 ($\alpha_{2000}$ = 
12$^{\rm h}$28$^{\rm m}$10\rlap{$^{\rm s}$}{.}9, $\delta_{2000}$ = 
+44$^\circ$06$'$47{\rlap{$''$}{.}\,7) was first diagnosed by its nonthermal
radio emission (Seaquist \& Bignell 1978).  Subsequent spectroscopic 
observations show narrow emission lines belonging to an HII region and broad 
emission lines of oxygen that are associated with dense supernova ejecta
(Kirshner \& Blair 1980).  The widths of these oxygen emission lines
suggest an expansion velocity of 3,500 km~s$^{-1}$.  The absence of 
broad [O II]$\lambda$3727 emission indicates an electron density greater 
than 10$^5$ cm$^{-3}$, and the [O III]$\lambda$4363 and $\lambda$5007 
line ratio implies an electron temperature of $\sim$40,000 K.
These properties strongly point to a case of young SNR dominated by
oxygen-rich ejecta of a supernova from a progenitor of mass $>$ 25
M$_\odot$.  The expansion of the high-density supernova ejecta into 
an HII region is responsible for the $\sim10^{39}$ erg~s$^{-1}$ X-ray 
luminosity (Blair, Kirshner, \& Winkler 1983).

The super-luminous SNR in NGC\,4449 has been recently studied by
Blair \& Fesen (1998) with high-resolution images taken with the
HST Faint Object Camera (FOC), and with high-S/N spectra taken with the
HST Faint Object Spectrograph and the 2.4m telescope at the MDM 
Observatory.  The HST FOC images show that the SNR is unresolved by 
the point spread function of the HST (Figure 1), placing an upper limit 
of 0\rlap{$''$}{.}\,028 (or 0.6 pc) on its diameter.  The new spectra
show broad wings of many emission lines indicating expansion velocities 
greater than 6,000 km~s$^{-1}$.  Blair and Fesen conclude that the
super-luminous SNR in NGC\,4449 is less than 100 years old, and that
its high luminosity is owed to its youth and expansion into a dense 
circumstellar environment. 

\begin{figure}[h]
\epsscale{0.4}
\plotone{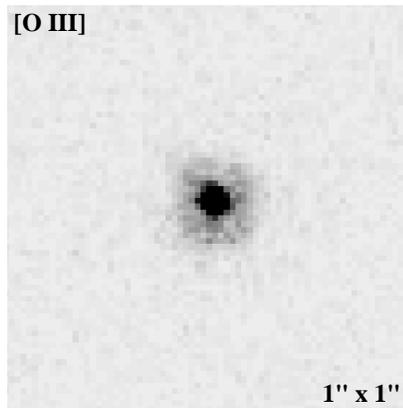}
\caption{HST FOC [O III]$\lambda$5007 image of the 
super-luminous SNR in NGC\,4449.  The image is unresolved and shows 
basically the point spread function of the HST.  The field of view 
is $1''\times1''$ (1$''$ = 24 pc).}
\end{figure}

\section{The Super-luminous SNR in NGC\,6946}

The super-luminous SNR in NGC\,6946 ($\alpha_{2000}$ = 
20$^{\rm h}$35$^{\rm m}$00\rlap{$^{\rm s}$}{.}75, $\delta_{2000}$ = 
+60$^\circ$11$'$30{\rlap{$''$}{.}\,6) was initially identified in an 
optical survey (Blair \& Fesen 1994).  It was immediately obvious that
this SNR is associated with an extremely luminous X-ray source.  With 
an X-ray luminosity of 2.8$\times$10$^{39}$ ergs~s$^{-1}$, this SNR
in NGC\,6946 has the highest X-ray luminosity among all known SNRs
(Schlegel 1994).  Radio observations show that this SNR is three times
as luminous as Cas A (Van Dyk et al.\ 1994).  Such high luminosities
are often associated with young SNRs; however, no high-velocity gas 
with expansion velocity greater than 600 km~s$^{-1}$ is detected
(Blair \& Fesen 1994).

\begin{figure}[h]
\epsscale{0.95}
\plotone{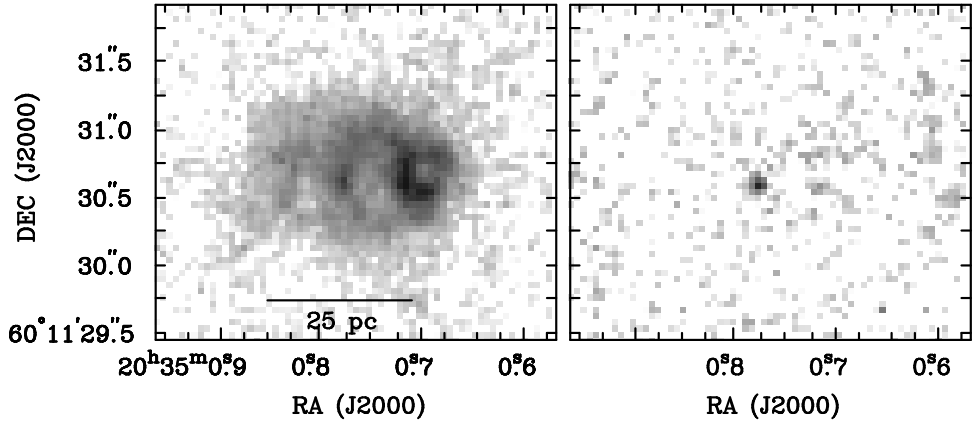}
\caption{HST WFPC2 images of the super-luminous SNR in NGC\,6946
in H$\alpha$ (left) and F439W blue continuum (right).  These images
are downloaded from the HST archive and plotted by Bryan Dunne.}
\end{figure}

HST Wide Field Planetary Camera 2 (WFPC2) images of the super-luminous
SNR in NGC\,6946 were recently reported by Blair, Fesen, \& Schlegel 
(1997).  The WFPC2 H$\alpha$ (F656N filter) image shows that this 
SNR is in an isolated environment and that the remnant has a 
multiple-loop morphology with an oblong outermost shell 20$\times$30 pc
in size (Figure 2).  Based on this morphology, Blair et al.\ suggested 
that the high X-ray luminosity is caused by colliding SNRs, with the 
small, bright loop ($\sim$8 pc in diameter) on the western part being 
a young SNR and the outermost shell being an older SNR.

\begin{figure}[h]
\plotone{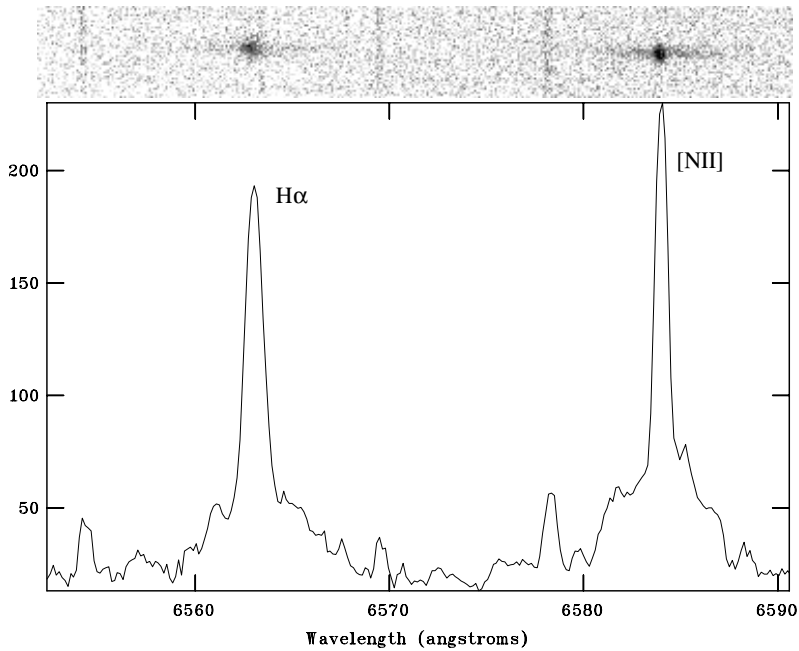}
\caption{KPNO 4m echelle spectra of the super-luminous SNR in NGC\,6946
in the region of H$\alpha$ and [N II] lines.  Both lines show a narrow
component superposed on a broad component.  The weak narrow lines are
the geocoronal H$\alpha$ and telluric OH lines.}
\end{figure}

To investigate the kinematic properties of the super-luminous SNR 
in NGC\,6946, we have obtained high-dispersion spectra using the
echelle spectrograph on the 4m telescope at Kitt Peak National
Observatory (Dunne, Gruendl, \& Chu 1999).  As shown in Figure 3, 
the nebular emission lines are resolved into a narrow component 
superposed on a broad component.  The velocity width and 
[N II]$\lambda$6583/H$\alpha$ line ratio can be measured separately 
for the broad component and the narrow component.  These measurements 
are listed in Table 2.

The broad component must originate in SNR shocked material, while
the narrow component consists of unshocked gas.  The base of the broad
component, the full width at zero intensity (FWZI), of the H$\alpha$
line extends over 450 km~s$^{-1}$, indicating an expansion 
velocity of at least 225 km~s$^{-1}$.  This expansion velocity is not
particularly high compared to those of SNRs in the Large Magellanic
Cloud (Chu \& Kennicutt 1988), and is usually associated with SNRs
that are $\sim$10$^4$ yrs old.   The lack of material expanding at
higher velocities is inconsistent with the identification of the
bright, small (8-pc across) loop as a young SNR.  The simple stellar
environment, shown by the WFPC2 blue continuum image in Figure 2,
also does not support a high supernova occurrence rate.  It is thus
not likely that the super-luminous SNR in NGC\,6946 consists of
colliding SNRs.

An effective clue to the nature of the super-luminous SNR in NGC\,6946 
is provided by the [N II]/H$\alpha$ ratios measured in the echelle
spectrum.  The [N II]/H$\alpha$ ratio is 0.8 in the narrow component
and 1.0 in the broad component.  An [N II]/H$\alpha$ ratio of 0.8 is 
hardly ever seen in an interstellar HII region, but is frequently seen
in ring nebulae around Wolf-Rayet (WR) stars or luminous blue variables
(LBVs).  The [N II] line is strong because the nebulae contain N-rich
material ejected by the central stars.  It is possible that the 
progenitor of the supernova in NGC\,6946's super-luminous SNR was a WR 
or LBV star, and the supernova ejecta interact with the dense 
circumstellar nebula and produce the high luminosity (Dunne et al.\ 1999).

The H$\alpha$ emission from the SNR shocked material can be determined
from the flux contained in the broad component of the velocity profile.
Assuming that the SNR shocked material is distributed in a shell whose
thickness is 1\% of its radius, Dunne et al. (1999) have derived a rms
electron density of $\sim$185 cm$^{-3}$, a mass of $\sim$1,300 M$_\odot$,
and a kinetic energy of $\sim$7$\times10^{50}$ ergs for the SNR shell.
This kinetic energy is somewhat high, but does not need an explosion
energy as high as those provided by GRBs.

\begin{table}
       \begin{tabular}{lcc}
      &      Broad    &    Narrow     \\
      &   component   &    Component  \\
\\
H$\alpha$~~~~~~FWHM~~(km~s$^{-1}$)    &  285   &  42  \\
$[$N II$]$~~~FWHM~~(km~s$^{-1}$)      &  250   &  25  \\
\\
H$\alpha$~~~~~~FWZI~~~~(km~s$^{-1}$)  &  450   &  ... \\
$[$N II$]$~~~FWZI~~~~(km~s$^{-1}$)    &  400   &  ... \\
\\
$[$N II$]$/H$\alpha$ flux ratio       &  1.0   &  0.8 \\
\\
H$\alpha$, ~~~broad/narrow flux ratio    &  1.5   &  ... \\
$[$N II$]$, broad/narrow flux ratio   &  1.8   &  ... \\
\\
        \end{tabular}
\caption{H$\alpha$ and [N II] Lines of the Super-luminous SNR in NGC\,6946}
\end{table}

\section{The Super-luminous SNR NGC\,5471B in M101}

NGC\,5471B is the B component of the giant HII region NGC\,5471 in M101.
The super-luminous SNR of NGC\,5471B was initially discovered by its
nonthermal radio emission and confirmed by its high [S II]/H$\alpha$
ratio (Skillman 1985).  High-dispersion echelle spectroscopic 
observations clearly show a broad emission line component at NGC\,5471B
(Figure 4).  Using the high-dispersion echelle spectra, Chu \& Kennicutt 
(1986) decomposed the H$\alpha$ velocity profile into a broad SNR 
component and a narrow HII region component, and derived a mass of 
6,500$\pm$3,000 M$_\odot$ and a kinetic energy of 
(2.5$\pm$1)$\times$10$^{50}$ ergs for the SNR.  ROSAT observations of 
NGC\,5471 show an X-ray source centered at the B component with a 
luminosity of 3$\times$10$^{38}$ ergs~s$^{-1}$ (Williams \& Chu 1995; 
Wang 1999).  From this X-ray luminosity, Wang (1999) derived a supernova
explosion energy of $\sim3\times10^{52}$ ergs, and suggested that 
NGC\,5471B is a ``hypernova remnant''.

\begin{figure}[t]
\epsscale{0.7}
\plotone{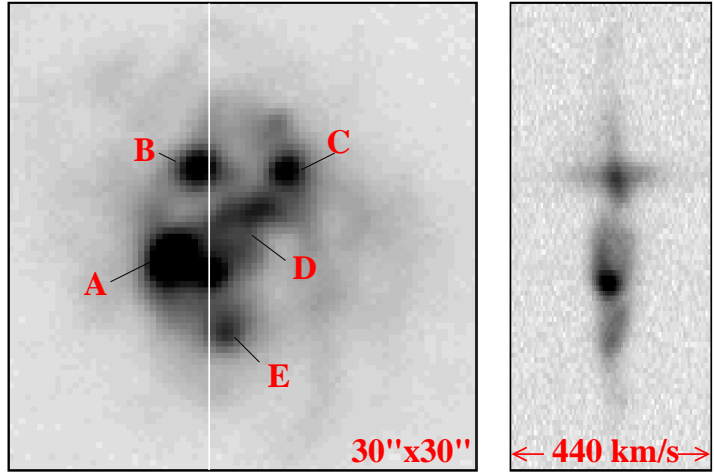}
\caption{KPNO 2.1m H$\alpha$ CCD image of NGC\,5471 (left) and KPNO 4m
 echelle H$\alpha$ line image (right), plotted with the same image
 scale.  The A--E components of NGC\,5471 and the echelle slit position
 are marked on the CCD image.  The field of view of this image is 
 $30''\times30''$ (1$''$ = 35 pc).  The echelle H$\alpha$ line image
 covers 440 km~s$^{-1}$ along the X-axis.  The broad emission component
 is clearly seen at the B component of NGC\,5471.}
\end{figure}

To study the physical nature of NGC\,5471B, we have obtained HST WFPC2
images in H$\alpha$ (F656N), [S II] (F673N), and continuum bands (F547M 
and F675W).  Comparisons between the H$\alpha$ and [S II] images of
NGC\,5471 show three shells with enhanced [S II]/H$\alpha$ ratios
(Figure 5).  The brightest of these [S II]-bright shells is the
super-luminous SNR in the B component.  

A closer examination of NGC\,5471B (Figure 6) shows that the 
super-luminous SNR is embedded in a very complex environment.
The H$\alpha$ image shows roughly a shell structure, with a bright
compact HII region at the southeast rim.  The blue continuum (F547M)
image shows individual supergiants, OB associations, and clusters.
The compact HII region on the shell rim is coincident with an OB
association/cluster.  The [S II] image shows the brightest [S II]
emission on the western side of the shell, where the underlying 
stellar emission is the lowest.  The areas  of enhanced 
[S II]/H$\alpha$ ratio extend beyond the southwest rim of the shell.

\begin{figure}[h]
\epsscale{0.97}
\plotone{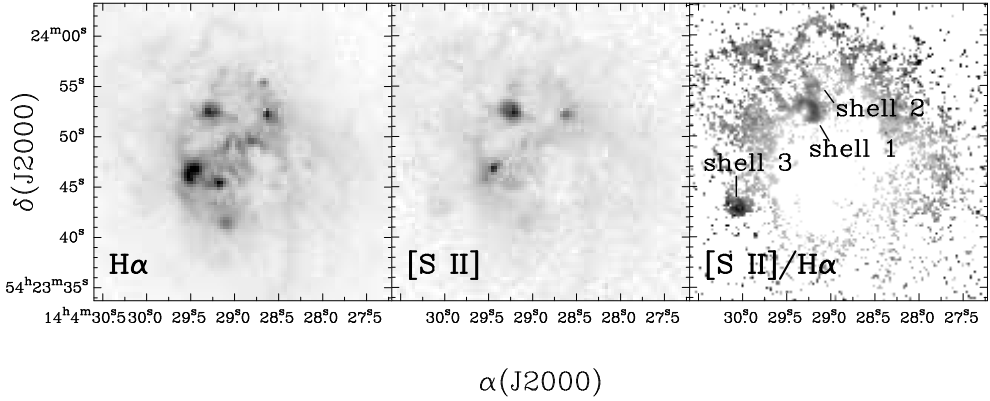}
\caption{HST WFPC2 images of NGC\,5471.  The left two images are
in H$\alpha$ and [S II] lines, respectively.  The right panel 
shows the [S II]/H$\alpha$ ratio map.  The three [S II]-enhanced 
shells are labeled.  Shell 1 is the super-luminous SNR in 
NGC\,5471B.}
\end{figure}

\begin{figure}
\epsscale{0.755}
\plotone{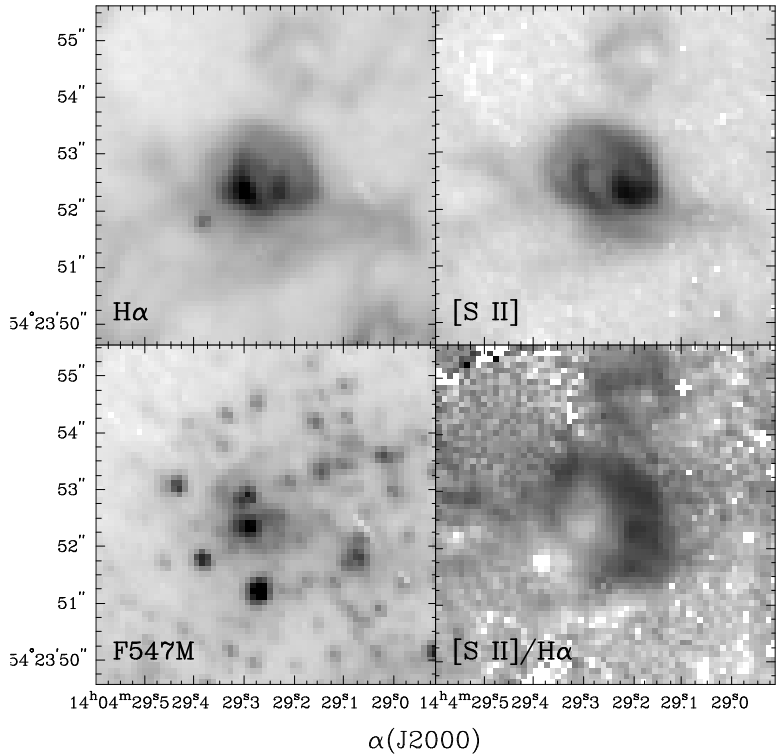}
\caption{HST WFPC2 images of NGC\,5471B in H$\alpha$ (F656N) and 
[S II] (F673N) lines and a blue continuum (F547M) band.  The
lower right panel is the [S II]/H$\alpha$ ratio map.  The H$\alpha$
image shows the ionized gas, [S II] image the shocked gas, and the 
F547M image the stars.  The brightest concentrations of stars are
OB associations or clusters.}
\end{figure}

To study the physical properties of the super-luminous SNR in 
NGC\,5471B, it is necessary to separate the SNR emission from 
the background HII region emission.  This can be achieved 
kinematically, using high-dispersion spectra.  We have obtained
new observations of NGC\,5471 with the echelle spectrograph
on the KPNO 4m telescope in 1999.  These new spectra are deeper 
and have higher S/N that those presented by Chu \& Kennicutt (1986).
The new echelle data are analyzed similarly.  As shown in Figure 7,
the H$\alpha$ velocity profile can be fitted by two Gaussian 
components, with the broad component corresponding to SNR emission
and the narrow component the background HII region emission.  
The broad component contributes to $\sim$70\% of the total flux.  
Its FWHM is 135$\pm$5 km~s$^{-1}$, but the faint wings extends over 
$\pm$210 km~s$^{-1}$.  

\begin{figure}[t]
\epsscale{0.45}
\plotone{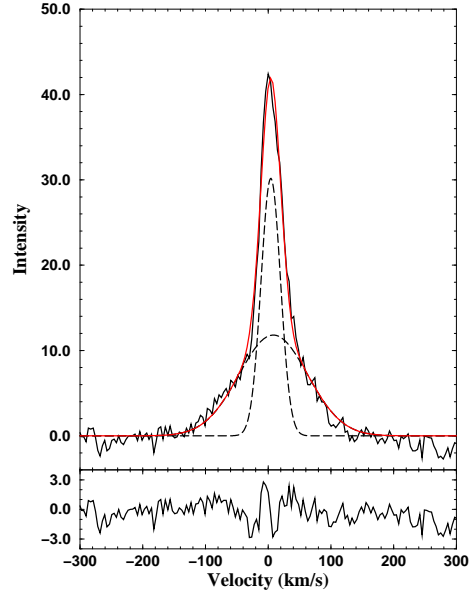}
\caption{H$\alpha$ line profile of NGC\,5471B.  The profile is fitted 
by two Gaussian components.  The dashed curves are the two components.
The bottom panel plots the residuals.}
\end{figure}

The H$\alpha$ luminosity of the SNR in NGC\,5471B can be determined
using the H$\alpha$ flux derived from the HST WFPC2 H$\alpha$ image, 
the broad/total flux ratio derived from the H$\alpha$ velocity 
profile, and the extinction derived spectroscopically by Kennicutt \& 
Garnett (1996).  The luminosity of the SNR so derived is 
1.6$\times$10$^{39}$ ergs~s$^{-1}$.  Assuming that the SNR shell has
a shell thickness 1\% of the shell radius, we derive a mass of 
1.8$\times$10$^4$ M$_\odot$.  Adopting an expansion velocity of 
210 km~s$^{-1}$ (one half of the full velocity extent), we derive
a shell kinetic energy of 8.2$\times$10$^{51}$ ergs.
This kinetic energy is more than one order of magnitude higher than
those seen in SNRs in the Magellanic Clouds (Williams et al.\ 1997, 1999).
If the SNR in NGC\,5471B is indeed a SNR caused by one single supernova
explosion, the shell kinetic energy implies that the supernova explosion 
energy must be 3--30 times higher, depending on whether Sedov's solution 
or Chevalier's (1974) model is adopted.  An explosion energy of 
2$\times$10$^{52}$ -- 2$\times$10$^{53}$ ergs is 1--2 orders of magnitude
higher than the canonical explosion energy of 10$^{51}$ ergs
(Jones et al. 1998).  

A crucial question to ask is whether the energetic, expanding shell
in NGC\,5471B is indeed energized by a single supernova with an
extraordinary explosion energy or a large number of stars over a long
period of time.  As the shell interior does not have a high concentration
of stars and the [S II]-bright part of the shell is particularly devoid
of stars, it is quite likely that the shell structure is a SNR, instead
of a  well-developed superbubble recently energized by an interior 
supernova.  Further discussion is given in Section 8.

\section{The Super-luminous SNR MF83 in M101}

The super-luminous SNR MF83 was cataloged by Matonick \& Fesen (1997).
H$\alpha$ and V-band images of MF83 and surrounding regions have been
obtained with the MDM 2.4m telescope by Eva Grebel, who has kindly made
these images available to us.  Figure 8 shows that MF83 is located 
between two spiral arms on the eastern part of M101.  The H$\alpha$ 
image in Figure 9 shows that this remnant does not have a clear shell
morphology.  The bright spots on the south rim of the remnant corresponds
to stars, which are better seen in the V-band image in Figure 9.

\begin{figure}[t]
\plotone{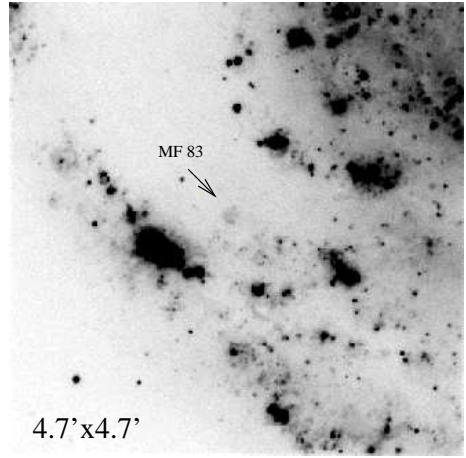}
\caption{MDM 2.4m H$\alpha$ image of MF83 and the eastern portion of M101.
(Photo credit: Eva Grebel)}
\end{figure}

\begin{figure}[t]
\epsscale{0.97}
\plotone{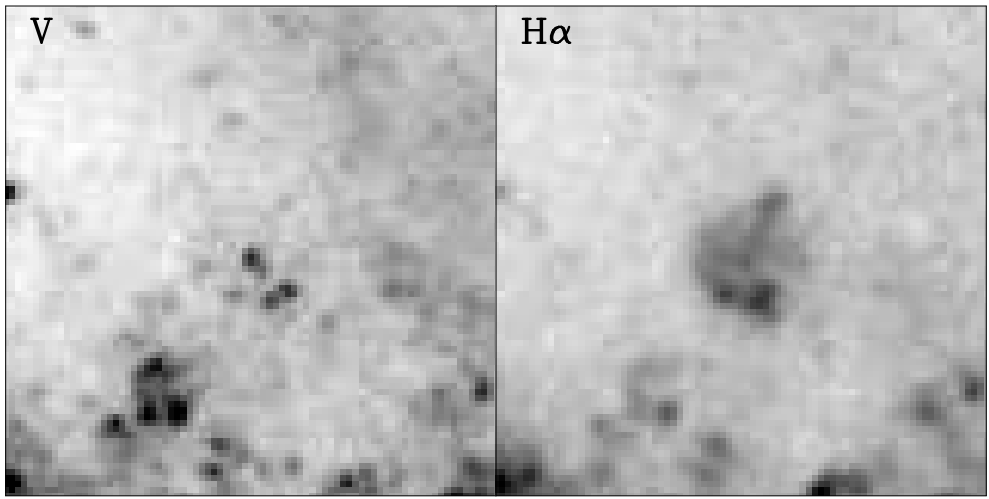}
\caption{MDM 2.4m images of MF83 in the V-band and the H$\alpha$ 
line.  The field of view is $45''\times45''$. (Photo credit: Eva Grebel)}
\end{figure}

We have obtained high-dispersion echelle spectra of MF83 with the 
KPNO 4m telescope in 1999.  The echelle observations detected MF83 
in H$\alpha$, [N II], and [S II] lines.  The H$\alpha$ line,
having the highest S/N ratio, is displayed in Figure 10.  The
line image shows clearly an expanding shell structure, with the extreme
velocities extending over $\sim$145 km~s$^{-1}$.  The integrated 
H$\alpha$ velocity profile of MF83, after subtracting the telluric OH
line, can be fitted with a Gaussian component with a FWHM of 82$\pm$14 
km~s$^{-1}$ (see Figure 11).  We adopt the full extent of the H$\alpha$
profile as twice the expansion velocity.  The expansion velocity is 
thus $\sim$70 km~s$^{-1}$.

\begin{figure}[h]
\epsscale{0.6}
\plotone{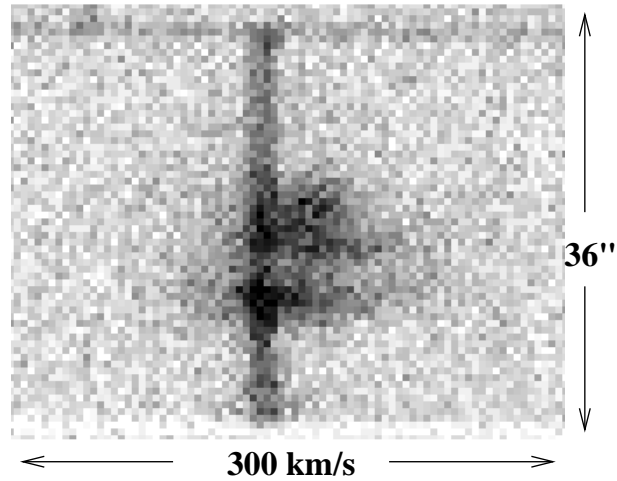}
\caption{KPNO 4m echelle image of the H$\alpha$ line.  The emission from
MF83 shows a position-velocity ellipse indicating an expanding-shell 
structure.  The narrow, constant component is a telluric OH line.}
\end{figure}

\begin{figure} [h]
\epsscale{0.45}
\plotone{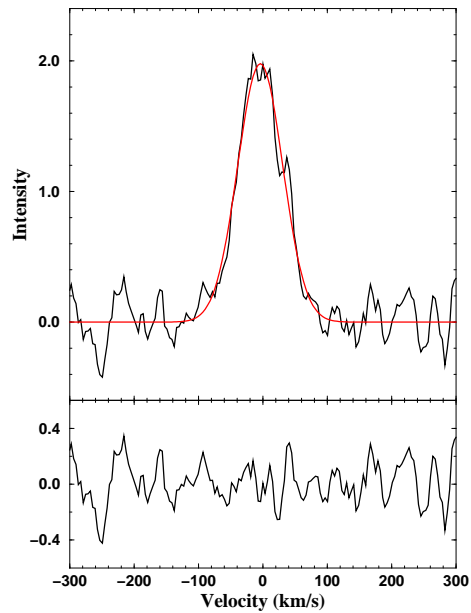}
\caption{Integrated H$\alpha$ velocity profile of MF83.  The superposed
telluric OH line has been removed.  The smooth curve is the Gaussian fit.
The residuals are plotted in the bottom panel.}
\end{figure}

We adopt the H$\alpha$ luminosity 1.7$\times$10$^{38}$ ergs~s$^{-1}$ 
from Matonick \& Fesen (1997) and a diameter of 267 pc for MF83.  
Assuming a shell geometry with a shell thickness of 1\% the shell
radius, we derive a shell mass of 5.5$\times$10$^4$ M$_\odot$ and
a kinetic energy of 3$\times$10$^{51}$ ergs.  If MF83 is formed
by one single supernova explosion, the explosion energy would
need to be 10$^{52}$ --10$^{53}$ ergs.  This is 1--2 orders of 
magnitude higher than the canonical explosion energy for a normal
supernova.

Is MF83 formed by one single supernova?  The answer is likely to be 
negative because (1) MF83 has a cluster at its center, (2) MF83's shell
size is comparable to those of superbubbles around OB associations or
clusters, and (3) MF83's shell expansion velocity is too low to 
produce X-ray emission.  This will be discussed further in the next
section.

\section{Are NGC\,5471B and MF83 Remnants Related to GRBs?}

It has been suggested that NGC\,5471B and MF83 are hypernova remnants
that require explosion energies comparable to the energies frequently 
associated with GRBs (Wang 1999).  We have analyzed these two remnants
using HST WFPC2 images, MDM 2.4m CCD images, and KPNO 4m echelle spectra.
The results of our analysis are summarized in Table 3.
While we find that the kinetic energies of these two remnants do require
explosion energies of 10$^{52}$ -- 10$^{53}$ ergs, we also find that they
are in complex environments which make it difficult to assess whether 
they were formed by single supernova explosions.

NGC\,5471B has nonthermal radio emission and soft X-ray emission that 
are characteristic of SNRs.  It has an OB association/cluster at its
rim, but its shell size is smaller than most known superbubbles.  
It is unlikely that the NGC\,5471B shell is a superbubble produced 
by this off-center OB association/cluster.  There is no obvious 
counter-evidence for the super-luminous SNR NGC\,5471B to be produced 
by a single explosion.  We consider that a single explosion is at least
as likely as multiple explosions.

MF83, on the other hand, is not a known nonthermal radio source, and
its ROSAT X-ray observations were too noisy to provide spectral information.
It has an OB association/cluster at its center and its shell size is 
comparable to those commonly seen in superbubbles.  MF83 was initially
identified as a SNR based on its high [S II]/H$\alpha$ ratio, but many
superbubbles are known to have high [S II]/H$\alpha$ ratios as well,
e.g., N185 and N186E in the Large Magellanic Cloud (Lasker 1977).
Its expansion velocity is on the high side of the known expansion 
velocities of superbubbles, but is similar to that of the aforementioned 
[S II]-bright superbubble N185 (70$\pm$10 km~s$^{-1}$, Rosado et al.\ 1982) 
and is too low to generate much X-ray emission.  We therefore conclude 
that there is no compelling evidence for MF83 to be produced by one single 
powerful hypernova explosion.

Finally, are NGC\,5471B and MF83 remnants of GRBs?  The explosion 
energies we have derived have large uncertainties; however, the
lower end of the energy range, a few times 10$^{52}$ ergs, is compatible
with some supernova energies determined recently by Branch (1999, this
volume).  We conclude that the association of these two super-luminous
SNRs with GRBs is at best remote.

\begin{table}
       \begin{tabular}{lccl}
Parameter          &    NGC\,5471B        &    MF83             &   ~~~Units      \\
\\ 
Size               &       60             &    270              &  ~~~pc          \\
V$_{\rm exp}$      &      210             &     70             & ~~~km~s$^{-1}$   \\
Mass               & 1.8$\times$10$^4$    & 5.5$\times$10$^4$   & ~~~M$_\odot$    \\
L(H$\alpha$)       & 1.6$\times$10$^{39}$ & 1.7$\times$10$^{38}$& ~~~ergs s$^{-1}$ \\
L(X)               & 1.7$\times$10$^{37}$ & 1$\times$10$^{37}$  & ~~~ergs s$^{-1}$ \\
Kinetic Energy     &  8$\times$10$^{51}$  & 3$\times$10$^{51}$  &  ~~~ergs     \\
Explosion Energy  & 2$\times$10$^{52}$ -- 2$\times$10$^{53}$ 
 & 10$^{52} - 10^{53}$ &  ~~~ergs \\
\\
Radio spectrum     &  nonthermal          &     ?      \\
X-ray spectrum     &   soft               &  ?     \\
OB association     &   off-center          &   at center  \\
Powerful Explosion &   very likely         &   less likely  \\
\\
        \end{tabular}
\caption{Physical Properties of NGC\,5471B and MF83}
\end{table}

\end{document}